\documentclass[aps,nofootinbib,11pt]{revtex4}
\usepackage{bm}% bold math

\begin{document}

%\preprint{}

\title{AdS$_3$ backgrounds from 10D effective action of heterotic string theory}

\author{Predrag Dominis Prester}%
 \email{pprester@phy.uniri.hr}
\affiliation{%
   Department of Physics, University of Rijeka\\
   Omladinska 14, HR-51000, Croatia}

% \date{\today}% It is always \today, today,
             %  but any date may be explicitly specified

\begin{abstract}
We present a method for calculating solutions and corresponding central charges for backgrounds 
with AdS$_3$ and $S^k$ factors in $\alpha'$-exact fashion from the \emph{full} tree-level low 
energy effective action of heterotic string theory. Three examples are explicitly presented: 
AdS$_3 \times S^3 \times T^4$, AdS$_3 \times S^2 \times S^1 \times T^4$ and 
AdS$_3 \times S^3 \times S^3 \times S^1$. Crucial property which enabled our analysis is 
vanishing of the Riemann tensor calculated from connection with "$\sigma$-model torsion". We 
show the following: (i) Chern-Simons terms are the only source of $\alpha'$-corrections not only in 
BPS, but also in non-BPS cases, suggesting a possible extension of general method of Kraus and 
Larsen, (ii) our results are in agreement with some conjectures on the form of the part of tree-level Lagrangian not connected to mixed Chern-Simons term by supersymmetry (and present in all 
supersymmetric string theories), (iii) new $\alpha'$-exact result for central charges in 
AdS$_3 \times S^3 \times S^3 \times S^1$ geometry. As a tool we used our generalization of 
Sen's $\mathcal{E}$-function formalism to AdS$_p$ with $p>2$, and paid special attention to
proper definition of asymptotic charges. 
\end{abstract}

\pacs{Valid PACS appear here}% PACS, the Physics and Astronomy
                             % Classification Scheme.
%\keywords{Suggested keywords}%Use showkeys class option if keyword
                              %display desired
\maketitle

\section{\label{sec:intro}Introduction}

String backgrounds with $AdS_3$ factor play important role in understanding of 
AdS$_{d+1}$/CFT$_d$ duality conjecture, as they provide examples in which we can perform 
calculations on both sides of the duality enabling in this way direct comparison. Some of these 
backgrounds are also connected to asymptotic near-horizon geometries of black strings. A specially important interplay of these two situations appears in microscopic calculations of entropy for 
extremal black holes. In string theory near-horizon geometries of such black holes typically contain 
AdS$_2 \times S^1$ factor which happens to be locally isometric to AdS$_3$. By calculating central charges $c_{L,R}$ of dual boundary CFT$_2$ (e.g., by using Cardy formula) one can calculate microcanonical entropies of corresponding extremal black holes in "microscopic" fashion. There are examples in which direct microscopic calculation of black hole entropy (i.e., without using AdS/CFT conjecture) is also possible. Agreement between microscopic and macroscopic (obtained from 
low-energy effective supergravity action and Wald formula) results for black hole entropies in all 
known examples (where calculation on both sides is applicable and possible) is one of the big achievements of string theory, as it shows that the theory provides correct statistical interpretation 
of black hole thermodynamics.

One motivation for studying $\alpha'$-corrections from the macroscopic side (using tree-level 
effective supergravity actions) is that this allows us to make more precise comparisons with 
microscopic ("stringy") calculations. In this way one can make precision tests of some of the most 
important results in string theory, like above mentioned statistical derivation of black hole 
thermodynamics). One can also turn the argument around, and use the (presupposed) agreement between macro- and micro- results to get some knowledge on the structure of the low-energy 
effective action. We shall use both of these strategies in this paper.

Now, in some cases $\alpha'$-exact results are known on the microscopic side, so it would be 
desirable to be able to do $\alpha'$-exact calculations also on the macroscopic side. On the first 
look, this appears as an impossible mission because tree-level low-energy effective actions of 
string theories are known only partially. In fact, they are known fully only up to $\alpha'^2$- 
(i.e., 6-derivative) order. Confronted with this obstacle, different strategies were investigated in
the literature. Some authors started with dimensionally reduced effective actions truncated to four 
derivatives ($R^2$-type actions), with the most popular candidates being $R^2$-type 
supersymmetric actions and (non-supersymmetric) Gauss-Bonnet type actions (for reviews see 
\cite{Sen:2007qy,Castro:2008ne,PDPlosinj}). Surprisingly, in some supersymmetric (BPS) cases 
these actions produce $\alpha'$-exact results for black hole entropies and/or central charges (with 
Gauss-Bonnet actions working only in $D=4$ dimensions). However, it was also shown that this is 
not the case in non-supersymmetric (non-BPS) cases \cite{Sen:2007qy,Cvitan:2007en} (the same 
can be shown for central charges in $D=5$, by simple extension of the method from 
\cite{Cvitan:2007en} to AdS$_3\times S^2$ geometry), from which follows that these truncated 
actions are intrinsically incomplete already at 4-derivative, i.e., $\alpha'^1$-, order. Another 
approach was to analyze 3-dimensional effective actions defined on asymptotic AdS$_3$ 
geometries. This was immensely fruitful direction, giving along a way a simple method for 
obtaining $\alpha'$-exact results for central charges. By showing that all relevant information is 
encoded in Chern-Simons terms, this method also provided explanation for the sufficiency of 
terms with at most four derivatives in effective actions 
\cite{Kraus:2005vz,Kraus:2005zm,David:2007ak,Kaura:2008us}. However, so far all constructions 
were relying heavily on the presence of supersymmetry in effective 3-dimensional actions.

These developments raise several questions. Two of them, which partly motivated our work 
presented here, are: (1) What about non-supersymmetric backgrounds, for which the methods 
from previous paragraph are not applicable? Is the situation really different for them? (2) What is 
the reason for the mentioned irrelevance of 6- and higher-derivative terms, viewed from the 
perspective of 10-dimensional string effective actions? In 3-dimensional language combination of supersymmetry and symmetries of AdS$_3$ space guaranty this, but no corresponding argument 
is known in 10-dimensional language. Well known example are 8-derivative terms multiplied by 
$\zeta(3)$ number (including famous $\zeta(3)RRRR$ terms) which are present in 10-dimensional 
tree-level low-energy actions of all string theories. In many known examples these terms should 
combine to give vanishing contribution to central charges (and extremal black hole entropies), but 
no mechanism which enforces this is known (exact structure of these terms is still not known).

Following our previous work \cite{Prester:2008iu}, we investigate these questions by starting from 
the \emph{full} 10-dimensional tree-level low-energy effective action of heterotic string theory, with 
the idea of taking into account all higher-derivative terms, and finding $\alpha'$-exact 
solutions (and corresponding central charges) for backgrounds containing factors of AdS$_3$,
spheres and tori. We explicitly present three examples - AdS$_3 \times S^3 \times T^4$, 
AdS$_3 \times S^2 \times S^1 \times T^4$ and AdS$_3 \times S^3 \times S^3 \times S^1$, but
method obviously extends to other cases. For first and second example, microscopic results for 
central charges are known, so we are able to make comparison with our macroscopic calculations. 
As for the third example, as far as we know the results for central charges are new.

Because the full tree-level action is only partially known, our strategy is to first take into account the
part of the effective action which is connected by supersymmetry with (gauge-gravity) mixed 
Chern-Simons term which we are able to solve directly, without any assumptions. We obtain
solutions for all charge-signatures, which include both BPS and non-BPS cases. Comparison of 
central charges, obtained in this way, with microscopic results (which are known for the first two 
examples) shows agreement in all cases, BPS \emph{and} non-BPS. Moreover, all higher-derivative 
terms except Chern-Simons term happen to be irrelevant in our examples, \emph{both} for BPS 
and non-BPS solutions. This provides new insight to question (1) above.

To completely answer question (2), one needs to take into account also the part of the action which 
is not connected by supersymmetry to mixed Chern-Simons term. This part starts at 8-derivative 
($\alpha'^3$-) order, and contains above mentioned $\zeta(3)RRRR$ terms. Now, the exact 
structure of this part is unknown, and only some terms (of 4-point and 5-point type) are completely 
known. \emph{We conjecture that this part of the action gives vanishing contribution in our 
calculations due to the fact that all our solutions satisfy a property}
\begin{equation}
\overline{R}_{MNPQ} = 0 \;,
\end{equation} 
where $\overline{R}$ is the "torsional" Riemann tensor in 10-dimensions, which is calculated from 
modified connection $\overline{\Gamma} = \Gamma - H/2$ ("$\sigma$-model torsion"). Of course, 
with this we conjecture that this part of the action has a particular form, which in fact was already conjectured previously in the literature ("weak form"of the conjecture was first proposed in 
\cite{Metsaev:1987zx}, and "strong form" in \cite{Kehagias:1997cq}). What is important here is that 
the most recent calculations \cite{Richards:2008sa} show that the known (i.e., 4-point and 5-point) 
8-derivative terms indeed are in accord with our conjecture. With this we have answered the 
question (2).

Here is the outline of the paper. In section \ref{sec:senef} we generalize Sen's entropy function
formalism to AdS$_p$ cases with $p\ge2$. This formalism, which we call $\mathcal{E}$-function
formalism, is for $p=3$ is equivalent to the so called $c$-extremization. Section \ref{sec:hetsol} is 
the central part of the paper. In subsection \ref{ssec:10Dha} we review what is known about the 
structure of 10-dimensional tree-level low-energy effective action of heterotic string theory and the 
strategy of dealing with mixed Chern-Simons term. In subsections \ref{ssec:s3}-\ref{ssec:s3s3} we 
present explicit solutions and corresponding central charges for three heterotic backgrounds: 
AdS$_3 \times S^3 \times T^4$, AdS$_3 \times S^2 \times S^1 \times T^4$ and 
AdS$_3 \times S^3 \times S^3 \times S^1$. Here we paid special attention to the proper
definition of asymptotic charges. In subsection \ref{ssec:typeII} we extend the results 
to corresponding backgrounds in type-II theories. Concluding remarks are left to section 
\ref{sec:concl}.

\section{\label{sec:senef}Sen's $\bm{\mathcal{E}}$-function formalism for AdS$_{\bm{p}}$}

\subsection{\label{ssec:senef3}Backgrounds with AdS$_{\bm{3}}$ factor}

Our goal is to analyse solutions of higher-derivative gravity theories with geometries given by 
products of some number of (maximally symmetric) spaces AdS$_p$ and $S^q$. In the special 
case of AdS$_2 \times S^{D-2}$ geometry, the procedure is developed in \cite{Sen:2005wa} and 
is known as Sen's entropy function formalism. As such geometries appear as near-horizon limit
of static extremal black holes, Sen's formalism gained a huge popularity as the simplest method 
for calculation of entropy of such black holes (for a review including a detailed list of references see
\cite{Sen:2007qy}). We are interested in generalization of Sen's formalism to geometries with
AdS$_p$, $p>2$, factors, in particular $p=3$ case\footnote{Somewhat different extension of Sen's
entropy function formalism to general AdS$_p$ geometries was developed in \cite{Garousi:2007zb} 
(and used in \cite{Hohenegger:2008du,Garousi:2008fz}).}.

Let us assume that we have some purely bosonic theory of gravity in $D$-dimensions which is 
manifestly diffeomorphism covariant, and, if there are gauge symmetries, also manifestly gauge 
covariant. We want to find solutions with the AdS$_3 \times S^{D-3}$ geometry which are 
manifestly symmetric under the full group of isometries, i.e., $SO(2,2) \times SO(D-2)$. In that 
case the only fields (field strengths for gauge fields) which are not forced to vanish by 
symmetries are those which can be composed of "elementary tensors", which are metric tensors 
and volume-forms of AdS$_3$ and $S^{D-3}$ spaces.

For clarity, let us focus on a theories with a local Lagrangian $\mathcal{L}$ and a field content 
consisting of $D$-dimensional metric $G_{\mu\nu}$, scalars $\phi^{(s)}$, and some number of 
forms $\chi$ corresponding to gauge (in such cases $\chi$ denotes field strength) and auxiliary 
fields.\footnote{Bosonic sectors of low energy effective actions of string theories fall in this class.} 
Then the only potentially non-vanishing fields are metric, scalars and $p$-forms with $p=3,D-3,$ 
or $D$, which are constrained to have the following form:
\begin{eqnarray} \label{gads3s3}
&& ds^2 = G_{\mu\nu} dx^\mu dx^\nu = v_A\,ds_A^2 + v_S\,ds_S^2 \;,\qquad  \phi^{(s)} = u_s
 \nonumber \\
&& \chi_3^{(i)} = h_i\,\epsilon_A \;, \qquad \chi_{D-3}^{(a)} = h_a\,\epsilon_S \;, \qquad 
\chi_{D}^{(\alpha)} = h_\alpha\,\epsilon_A \wedge \epsilon_S
\end{eqnarray}
$ds_A^2$ ($ds_S^2$) and $\epsilon_A$ ($\epsilon_S$) denote metric and volume-form of 
AdS$_3$ ($S^{D-3}$) space with unit radius. This means that $v_A$ ($v_S$) is the squared radius 
of AdS$_3$ ($S^{D-3}$) appearing in the physical geometry. For convenience (and to make it as 
close to Sen's procedure as possible) we choose the coordinates in AdS$_3$ space such that the determinant of AdS$_3$ metric with unit radius is equal to $(-1)$. $v_A$, $v_S$, $u_s$ and $h's$ 
are constants. If $\chi_3^{(i)}$ is a gauge field strength, then we denote $h_i = e_i$. If 
$\chi_{D-3}^{(a)}$ is a gauge field strength, then $p_a = h_a$ is the magnetic charge (in some 
particular normalization). The rest of $h's$ are variables corresponding to auxiliary fields, which 
should be determined, together with $v_A$, $v_S$ and $u_s$, by solving the equations of motion.

If we define the function $f$ by
\begin{equation} \label{fads3}
f(\vec{v},\vec{u},\vec{h};\vec{e},\vec{p}) = \oint_{S^{D-3}} \sqrt{-G} \, \mathcal{L} \;,
\end{equation}
where we use the AdS$_3 \times S^{D-3}$ Ansatz (\ref{gads3s3}), then solving equations of 
motion is equivalent to extremization of the function $f$ keeping $\vec{e}$ and $\vec{p}$ fixed, i.e., 
to solving the algebraic system
\begin{equation} \label{eom}
0 = \frac{\partial f}{\partial \vec{v}} = \frac{\partial f}{\partial \vec{u}}
 = \frac{\partial f}{\partial \vec{h}} 
\end{equation}
It is more common to express results not in terms of electric fields $\vec{e}$ but in terms of 
electric charges $\vec{q}$ which are given by
\begin{equation} \label{qads3}
\vec{q} = \frac{\partial f}{\partial \vec{e}} \;.
\end{equation}
This transition goes through Legandre transformation, by introducing $\mathcal{E}$-function 
defined by
\begin{equation} \label{efads3}
\mathcal{E}(\vec{v},\vec{u},\vec{h},\vec{e};\vec{q},\vec{p})
 =  6\pi ( \vec{q} \cdot \vec{e} - f ) \;.
\end{equation}
Extremization of $\mathcal{E}$-function over variables $\vec{v},\vec{u},\vec{h},\vec{e}$
\begin{equation} \label{Eeom}
0 = \frac{\partial \mathcal{E}}{\partial \vec{v}}
 = \frac{\partial \mathcal{E}}{\partial \vec{u}}
 = \frac{\partial \mathcal{E}}{\partial \vec{h}}
 = \frac{\partial \mathcal{E}}{\partial \vec{e}} 
\end{equation}
then obviously gives (\ref{eom}) and (\ref{qads3}). Solving (\ref{Eeom}) one gets solutions for 
variables as functions of electric and magnetic charges $\vec{q}$ and $\vec{p}$. The value of 
$\mathcal{E}$-function at the extremum gives the central charge $c$ of the dual CFT living on 
the boundary of AdS$_3$ space
\begin{equation} \label{cads3}
c(\vec{q},\vec{p}) = \mathcal{E}(\vec{v}_0,\vec{u}_0,\vec{h}_0,\vec{e}_0;\vec{q},\vec{p}) \,,
 \quad (\vec{v}_0,\vec{u}_0,\vec{h}_0,\vec{e}_0\;\text{ satisfy (\ref{Eeom})})
\end{equation}

It is almost obvious that the above generalization of Sen's formalism to AdS$_3$ is equivalent to 
the so-called $c$-extremization method developed by Kraus and Larsen 
\cite{Kraus:2005vz,Kraus:2005zm} (for a nice review see \cite{Kraus:2006wn}). To see this, let us 
first consider a case where there are no electrically charged gauge fields. Then it is easy to 
check that our $\mathcal{E}$-function is equal to the $c$-function of Kraus and Larsen. 

Let us now assume that there are also $n$ electrically charged gauge fields, which 3-form field 
strengths $F_3^{(i)}=d\,A_2^{(i)}$ are constrained by (\ref{gads3s3}) to have the form
\begin{equation} \label{f3e}
F_3^{(i)} = e_i \,\epsilon_A \;, \qquad i=1,\ldots,n
\end{equation}
The idea is to pass to the dual magnetic description by making Poincare duality transformation.
We introduce $n$ additional $(D-4)$-form gauge fields $C_{D-4}^{(i)}$ with $(D-3)$-form gauge 
field strengths  $K_{D-3}^{(i)} = d\,C_{D-4}^{(i)}$, and define a new Lagrangian density
\begin{equation} \label{tlag}
\widetilde{\mathcal{L}} = \mathcal{L} - \frac{1}{3!\,(D-3)!\sqrt{-G}} \sum_{i=1}^n
 \varepsilon^{\mu_1\cdots\mu_D} F_{\mu_1\mu_2\mu_3}^{(i)} K_{\mu_4\cdots\mu_D}^{(i)}
\end{equation}
where totally antisymmetric tensor density $\varepsilon^{\mu_1\cdots\mu_D}$ by definition
receives values $\pm1,0$. If we now treat 3-forms $F_3^{(i)}$ as \emph{auxiliary} fields (instead 
of gauge field strengths), Lagrangian $\widetilde{\mathcal{L}}$ leads to the equations of motion 
which are equivalent to those obtained from $\mathcal{L}$ (Euler-Lagrange equation for 
$C_{D-4}^{(i)}$ simply says that $F_3^{(i)}$ is closed, so locally 
$F_3^{(i)} = d\,A_2^{(i)}$, i.e., $F_3^{(i)}$ is a gauge field strength).

For AdS$_3 \times S^{D-3}$ solutions $F_3^{(i)}$ and $K_{D-3}^{(i)}$ are constrained by 
(\ref{gads3s3}) to have the form
\begin{equation} \label{ads3KF}
F_3^{(i)} = e_i \,\epsilon_A \,, \qquad 
K_{D-3}^{(i)} = \frac{\widetilde{p}_i}{\Omega_{D-3}} \epsilon_{D-3}
\end{equation}
where $\Omega_{D-3}$ is a volume of the $(D-3)$-sphere with unit radius, and $\widetilde{p}_i$
are magnetic charges. We emphasize again that $F_3^{(i)}$ are now auxiliary fields which should 
be treated as other auxiliary 3-form fields $\chi_3$ in (\ref{gads3s3}) (if such exist). The new 
Lagrangian $\widetilde{\mathcal{L}}$ does not contain any electrically charged gauge fields and 
so can be used to perform $\mathcal{E}$-function formalism to obtain AdS$_3 \times S^{D-3}$ 
solutions and central charge of dual CFT. For this we define
\begin{equation} \label{tfads3}
\widetilde{f}(\vec{v},\vec{u},\vec{h},\vec{e};\vec{\widetilde{p}},\vec{p})
 = \oint_{S^{D-3}} \sqrt{-G} \, \widetilde{\mathcal{L}} \;,
\end{equation}
from which we obtain $\mathcal{E}$-function
\begin{equation} \label{tefads3}
\widetilde{\mathcal{E}}(\vec{v},\vec{u},\vec{h},\vec{e};\vec{\widetilde{p}},\vec{p})
 =  -6\pi \widetilde{f}
 = 6\pi \left( \vec{\widetilde{p}} \cdot \vec{e} - f \right) \;.
\end{equation}
The second equality follows from (\ref{tlag}) and (\ref{tfads3}). It is now obvious that if we 
make identification
\begin{equation} \label{tpe}
\widetilde{p}_i = q_i
\end{equation}
then $\mathcal{E}$-function $\widetilde{\mathcal{E}}$ (\ref{tefads3}) is equal to $\mathcal{E}$ 
from (\ref{efads3}), and so it is irrelevant which one is used for finding solutions and central 
charges. As $\widetilde{\mathcal{E}}$ is equivalent to $c$-function of Kraus and Larsen, this 
completes the proof.

$\mathcal{E}$-function formalism is easily generalised to 
AdS$_3 \times S^{k_1} \times \ldots \times S^{k_n}$ geometries. Though the generalisation is
straightforward, expressions are quite cumbersome and so we shall not write them for general
case. Instead, we present in Section \ref{ssec:s3s3} an explicit example with 
AdS$_3 \times S^3 \times S^3$ geometry.

\subsection{\label{ssec:gCS}Generalisation to theories with Chern-Simons terms}

Effective actions of string theories typically contain Chern-Simons terms which are not 
manifestly gauge or diffeomorphism covariant. This prevents direct application of 
$\mathcal{E}$-function method. Unfortunately, there is no general recipe for dealing with
Chern-Simons terms. Here, we shall restrict ourselves to terms of the type
\begin{equation} \label{mixCS}
S_{\rm CS} = \int T^{(D-3)} \wedge \Omega^{(\rm L3)}
\end{equation}
where $T^{(D-3)}$ is some manifestly (gauge and diff) covariant $(D-3)$-form, and 
$\Omega^{\rm L3}$ is 3-dimensional gravitational Chern-Simons term defined with
\begin{equation} \label{gcsn}
\Omega^{(\rm L3)}_{\mu\nu\rho} = \frac{1}{2} \, \Gamma^{\sigma}_{\mu\tau} \,
 \partial_\nu \Gamma^{\tau}_{\rho\sigma} \, 
 + \, \frac{1}{3} \, \Gamma^{\sigma}_{\mu\tau} \, \Gamma^{\tau}_{\nu\xi} \,
 \Gamma^{\xi}_{\rho\sigma} \;\; \mbox{(antisym. in $\mu,\nu,\rho$)}
\end{equation}
Applying now AdS$_3 \times S^{k_1} \times \ldots \times S^{k_n}$ Ansatz to $T^{(D-3)}$, one 
has to properly define the term $\varepsilon^{abc} \Omega^{(\rm L3)}_{abc}$. Obviously, this 
term can a priori live only on AdS$_3$, $S^3$, or $S^2 \times S^1$, but it is straightforward to 
check that it gives vanishing contribution for $S^3$ and $S^2 \times S^1$ (after integration over 
the respective volumes, present in definition of $\mathcal{E}$-function). This leaves us only with
AdS$_3$ case, for which we add the following rule to $\mathcal{E}$-function formalism:

\begin{itemize}
\item On AdS$_3$ one takes $\varepsilon^{abc} \Omega^{(\rm L3)}_{abc} = \pm 4$, where plus 
(minus) sign is used for left (right) central charge $c_L$ ($c_R$).
\end{itemize}

To prove this, we simply refer to $c$-extremization method \cite{Kraus:2006wn}. 
From it follows that when we apply $\mathcal{E}$-function formalism by neglecting all terms of 
the type (\ref{mixCS}), Eq. (\ref{cads3}) is giving us average central charge $(c_L + c_R)/2$. 
Gravitational Chern-Simons terms introduce diffeomorphism anomaly in dual CFT and generate 
a difference between $c_L$ and $c_R$ 
\begin{equation} \label{difano}
c_L - c_R = 48 \pi \beta \;,
\end{equation}
where $\beta$ is (in $\mathcal{E}$-function formalism language) a factor appearing in a 
contribution of CS term (\ref{mixCS}) to function $f$ defined in (\ref{fads3}). More precisely,
\begin{equation} \label{fCS}
f_{\rm CS} = \beta \, \varepsilon^{abc} \Omega^{(\rm L3)}_{abc}
\end{equation}
By using this in (\ref{efads3}) and (\ref{cads3}), comparison with (\ref{difano}) leads directly to 
our simple rule, which completes the proof. 

In string theory the $D$-dimensional effective theory is frequently obtained by Kaluza-Klein 
compactification on one or more circles $S^1$. In such cases, sometimes it is more practical
to calculate $\mathcal{E}$-function before compactification (in higher-dimensional space). In
such cases, it can happen that we need to calculate 
$\varepsilon^{abc} \Omega^{(\rm L3)}_{abc}$ on $S^2 \times S^1$ space on which metric is 
not factorized but instead has Kaluza-Klein form
\begin{equation} \label{3KK2}
ds^2 = g_{ab}(x) dx^a dx^b
 = \phi(x) \left[ g_{mn}(x) dx^m dx^n + \left(dy + 2A_m(x) dx^m\right)^2 \right] \,, 
\end{equation}
where $1\le a,b\le3$ and $1\le m,n\le2$. Following \cite{Guralnik:2003we,Sahoo:2006vz} we 
take
\begin{equation} \label{CS3cov}
\varepsilon^{abc} \Omega^{(\rm L3)}_{abc} = 
 \frac{1}{2} \varepsilon^{mn} \left[ R^{(2)} F_{mn}
  + 4 g^{m'p'} g^{q'q} F_{mm'} F_{p'q'} F_{qn} \right] \;,
\end{equation}
where $F_{mn}=\partial_m A_n - \partial_n A_m$ and $R^{(2)}$ is a Ricci scalar calculated from
$g_{mn}$. Then (\ref{CS3cov}) gives us the desired manifestly covariant form (in the reduced 
2-dimensional space) for the Chern-Simons term. This logic was originally applied in analyses
of extremal heterotic black holes (AdS$_2$ case) in \cite{Sahoo:2006pm,Cvitan:2007hu}. We 
shall use it here in Sec. \ref{ssec:s2}.

\subsection{\label{ssec:gadsD}Generalisation to AdS$\bm{_{d+1}}$}

It is natural to contemplate the extension of $\mathcal{E}$-function formalism to the backgrounds
with general AdS$_{d+1}$ factor, where $d\ge1$. It is obvious that we can straightforwardly 
generalize the formal procedure given in Eqs. (\ref{gads3s3}-\ref{cads3}), where now we define 
$\mathcal{E}_d$-function 
\begin{equation} \label{efadsd}
\mathcal{E}_d(\vec{v},\vec{u},\vec{h},\vec{e};\vec{q},\vec{p})
 =  \pi^{[(d+1)/2]} ( \vec{q} \cdot \vec{e} - f ) \;.
\end{equation}
$[x]$ denotes integer part of $x$. Extremal value of $\mathcal{E}_d$-function we denote by 
$c_d$
\begin{equation} \label{cadsd}
c_d(\vec{q},\vec{p}) = \mathcal{E}_d(\vec{v}_0,\vec{u}_0,\vec{h}_0,\vec{e}_0;\vec{q},\vec{p})
 \quad (\vec{v}_0,\vec{u}_0,\vec{h}_0,\vec{e}_0\;\text{ satisfy (\ref{Eeom})})
\end{equation}
The question is what is the meaning of $c_d$? As is known, $2c_1$ is the number of ground
states in dual $CFT_1$ (i.e., conformal quantum mechanics) and $6c_2$ is the central 
charge of dual $CFT_2$. For $d=2n$ even, one generalization of $d=2$ case is in fact known - it
gives the coefficient in trace anomaly of A-type in the dual CFT$_{2n}$ \cite{Henningson:1998gx}. More precisely, the trace anomaly is \cite{Imbimbo:1999bj}
\begin{equation} \label{wanom}
\mathcal{A}_{2n}(x) = \frac{c_{2n}}{(4\pi)^n (n!)^2} E_{n}(x) + \ldots \;,
\end{equation}
where $E_n$ denotes the Euler density in $d=2n$ dimensions, and dots $\ldots$ denote B-type
(conformally invariant) contribution. 

From now on we shall deal with $d=2$ case exclusively.

\section{\label{sec:hetsol}$\bm{\alpha'}$-exact solutions in heterotic string theory}

\subsection{\label{ssec:10Dha}Effective action and 10D SUSY}

We are interested here in bosonic solutions with AdS$_3$ factors of the tree-level heterotic 
low-energy effective action in $D$-dimensions, with $10-D$ dimensions compactified on a 
torus $T^{10-D}$. We restrict ourselves to the most simple case in which torus is flat and all
Kaluza-Klein 1-form gauge fields are uncharged (vanishing). In addition, 10-dimensional 
($SO(32)$ or $E_8 \times E_8$) Yang-Mills (1-form) field is also taken to vanish \footnote{In
fact, those are not important restrictions. Eventually, we can use $O(26-D,10-D)$ T-duality of 
tree-level heterotic theory to obtain from our results central charges in general case.}. It follows 
that the only non-vanishing fields present in this sector are metric $G_{MN}$, Kalb-Ramond 
2-form gauge field $B_{MN}$, and dilaton $\Phi$. As discussed in detail in \cite{Prester:2008iu}, 
effective Lagrangian can be decomposed in the following way
\begin{equation} \label{10dl}
\mathcal{L}^{\rm (H)} = \mathcal{L}_{01} + \Delta \mathcal{L}_{\rm CS} + \mathcal{L}_{\rm other} \;.
\end{equation}
The first term in (\ref{10dl}), explicitly written, is
\begin{equation} \label{l100}
\mathcal{L}_{01} = \frac{e^{-2\Phi}}{16\pi G_D} \left[ R +
 4 (\partial \Phi)^2 - \frac{1}{12} \overline{H}_{MNP} \overline{H}^{MNP} \right] \;,
\end{equation}
where $G_{D}$ is $D$-dimensional Newton constant.
3-form gauge field strength is not closed, but instead given by
\begin{equation} \label{hbcs}
\overline{H}_{MNP} = \partial_M B_{NP} + \partial_N B_{PM} + \partial_P B_{MN}
  - 3 \alpha' \overline{\Omega}_{MNP} \;,
\end{equation}
where $\overline{\Omega}_{MNP}$ is the gravitational Chern-Simons form
\begin{equation} \label{gcs}
\overline{\Omega}_{MNP} = \frac{1}{2} \, \overline{\Gamma}^{R}_{\;\; MQ} \,
 \partial_N \overline{\Gamma}^{Q}_{\;\; PR} \, 
 + \, \frac{1}{3} \, \overline{\Gamma}^{R}_{\;\; MQ} \, \overline{\Gamma}^{Q}_{\;\; NS} \,
 \overline{\Gamma}^{S}_{\;\; PR} \;\; \mbox{(antisym. in $M,N,P$)}
\end{equation}
Bar on the geometric object means that it is calculated using the modified connection
\begin{equation} \label{modcon}
\overline{\Gamma}^{P}_{\;\; MN} = \Gamma^{P}_{MN} - \frac{1}{2} \overline{H}^{P}_{\;\; MN}
\end{equation}
in which 3-form $\overline{H}$ plays the role of a torsion.

A presence of $\alpha'$-correction in $\mathcal{L}_{01}$, induced by Chern-Simons term 
through (\ref{hbcs}), breaks the supersymmetry. As shown in \cite{Bergshoeff:1989de}, one can 
retrieve supersymmetry (which is $\mathcal{N}=1$ in $D=10$) by introducing the term 
$\Delta \mathcal{L}_{\rm CS}$ in Lagrangian (\ref{10dl}). As this ''supersymmetrization of 
Chern-Simons term'' is an on-shell construction, it follows that $\Delta \mathcal{L}_{\rm CS}$ 
contains tower of higher-derivative terms, i.e., (probably infinite) expansion in $\alpha'$, starting 
at $\alpha'1$ (4-derivative) order. In \cite{Bergshoeff:1989de} it was shown that there is a field 
redefinition scheme in which $\Delta \mathcal{L}_{\rm CS}$ can be written by \emph{purely} using 
modified Riemman tensor $\overline{R}_{MNPQ}$ (calculated from modified connection 
(\ref{modcon})) 
\begin{equation} \label{modrie}
\overline{R}^{M}_{\;\;\; NPQ} = R^{M}_{\;\;\; NPQ} + \nabla_{[P} \overline{H}^{M}_{\;\;\;\: Q]N}
 - \frac{1}{2} \overline{H}^{M}_{\;\;\; R[P} \overline{H}^{R}_{\;\;\; Q]N} \;,
\end{equation} 
and the metric tensor (needed just to contract indices). As in \cite{Prester:2008iu}, we shall use 
this property to show that $\Delta \mathcal{L}_{\rm CS}$ is giving \emph{vanishing contribution} 
to solutions we construct in this paper.

A much less is known about $\mathcal{L}_{\rm other}$ part of the tree-level effective action. It 
is known that it contains tower of terms starting at 8-derivative ($\alpha'^3$) order, with the
notorious $R^4$-type terms multiplied by $\zeta(3)$ transcendental number. By now, only 
$R^4$, $R^3H^2$ and $RH^2(\nabla H)^2$ terms\footnote{Notation $R^kH^m$ 
denotes all monomials which can be written by multiplying and contracting $k$ Riemann tensors
and $m$ 3-form strengths $H$.} are fully known, and their structure is consistent with the 
conjecture that $\mathcal{L}_{\rm other}$ can be written by purely using modified Riemann tensor 
$\overline{R}_{MNPQ}$ \cite{Richards:2008sa}. If this conjecture is true (at least in a weaker form,
see section \ref{sec:concl}), then $\mathcal{L}_{\rm other}$ would also be irrelevant for our results 
(for the same reason as $\Delta \mathcal{L}_{\rm CS}$). We postpone further discussion to section 
\ref{sec:concl}. At the moment we shall simply ignore this term.

We apply the following strategy \cite{Prester:2008iu}. First we ignore the terms 
$\Delta \mathcal{L}_{\rm CS}$ and $\Delta \mathcal{L}_{\rm CS}$ in the tree-level heterotic 
effective action (\ref{10dl}), which leaves us with simpler reduced Lagrangian
\begin{equation} \label{redl}
\mathcal{L}_{\rm red} = \mathcal{L}_{01} \;.
\end{equation}
This (non-supersymmetric) action still has nontrivial $\alpha'$-corrections due to presence of 
gravitational Chern-Simons term in (\ref{hbcs}). Then we show that all of our \emph{exact} 
solutions (obtained from $\mathcal{L}_{\rm red}$) satisfy the condition
\begin{equation} \label{modrie0}
\overline{R}_{MNPQ} = 0 \;.
\end{equation}
Due to the structure of $\Delta \mathcal{L}_{\rm CS}$ term mentioned above, from (\ref{modrie0})
follows immediately that such solutions are also solutions of the supersymmetric action
\begin{equation} \label{susyl}
\mathcal{L}_{\rm susy CS} = \mathcal{L}_{01} + \Delta \mathcal{L}_{\rm CS} \;.
\end{equation}

Now, to use $\mathcal{E}$-function method presented in Section \ref{sec:senef} we have to find
a way to treat Chern-Simons term. We now show how to do this if it appears in the action as in 
(\ref{mixCS}). This can be achieved through the generalization of the method from 
\cite{Prester:2008iu} to general $D$, by the particular Poincare duality transformation (\ref{tlag}) in 
which one takes $F_3=dB_2$ (where $B_2$ is the 2-form Kalb-Ramond field $B_{MN}$ from 
(\ref{hbcs})). As $\overline{H}_{MNP}$ becomes now an auxiliary field ((\ref{hbcs}) does not apply), 
the only appearance of Chern-Simons term is of the form (\ref{mixCS}). 

Let us present this in more detail. The dual, classically equivalent, Lagrangian 
$\widetilde{\mathcal{L}}$ is defined by\footnote{Similar dual formulations are known for some
time, see, e.g., \cite{Nishino:1991sr}.}
\begin{equation} \label{ldual}
\widetilde{\mathcal{L}}^{\rm (H)} = \mathcal{L}^{\rm (H)} - \frac{3!}{(24\pi)^2 (D-3)!\sqrt{-G}} \,
 \varepsilon^{M_1\cdots M_D} 
 \left(\overline{H}_{M_1 M_2 M_3} + 3 \alpha' \overline{\Omega}_{M_1 M_2 M_3}  \right) 
 K_{M_4\cdots M_D}
\end{equation}
where it is understood that $(D-3)$-form $K$ is exact, i.e., $K=dC$, and 3-form $\overline{H}$ is 
treated as an auxiliary field.
Using (\ref{10dl}) we have
\begin{equation} \label{10dtl}
\widetilde{\mathcal{L}}^{\rm (H)} = \widetilde{\mathcal{L}}_0 + \widetilde{\mathcal{L}}_{\rm CS}
 + \Delta \mathcal{L}_{\rm CS} + \mathcal{L}_{\rm other} \;,
\end{equation}
where
\begin{eqnarray} \label{10dtl0}
\widetilde{\mathcal{L}}_0 &=& \mathcal{L}_0 - \frac{3!}{(24\pi)^2 (D-3)!\sqrt{-G}} \,
 \varepsilon^{M_1\cdots M_D} \overline{H}_{M_1 M_2 M_3} K_{M_4\cdots M_D} \\
\widetilde{\mathcal{L}}_{\rm CS} &=&
 - \frac{\alpha'}{32\pi^2 (D-3)!\sqrt{-G}} \, \varepsilon^{M_1\cdots M_D} 
 \overline{\Omega}_{M_1 M_2 M_3} K_{M_4\cdots M_D} \;.
\label{10dlcs}
\end{eqnarray}
In (\ref{10dtl}) and (\ref{10dtl0}) terms without tilde ($\mathcal{L}_0$, 
$\Delta \mathcal{L}_{\rm CS}$ and $\mathcal{L}_{\rm other}$) are the same as before with the 
important exception that in the dual description $H_{MNP}$ is now treated as an auxiliary field (so 
in the dual description we should forget the relation (\ref{hbcs})).

Let us pause for a moment to explain the terms in (\ref{10dtl}). $\widetilde{\mathcal{L}}_0$ is the
lowest ($\alpha'^0$-) order term in $\alpha'$-expansion. $\widetilde{\mathcal{L}}_{\rm CS}$ is the
mixed Chern-Simons term (constructed from torsional connection (\ref{modcon})) and in fact the
only term in the whole dualized heterotic effective action containing Chern-Simons term, and so
the only term which is not manifestly diffeomorphism-covariant. It is purely of $\alpha'^1$- 
(4-derivative) order. As mentioned before, $\Delta \mathcal{L}_{\rm CS}$ represents (probably 
infinite) tower of terms, starting at $\alpha'^1$-order, which are connected with Chern-Simons 
term by supersymmetry, and $\mathcal{L}_{\rm other}$ is "the rest", consisting of tower of terms 
starting at $\alpha'^3$- (8-derivative) order. What is important is that in the dual description, due 
to the auxiliary nature of 3-form $\overline{H}$, these terms are now free of Chern-Simons 
term (before dualization they have contained it implicitly because of (\ref{hbcs})).

Though we have extracted Chern-Simons term out, we are still not completely satisfied because
$\overline{\Omega}_{MNP}$ appearing in (\ref{10dlcs}) is, due to (\ref{modcon}), not the 
normal gravitational Chern-Simons term $\Omega_{MNP}$ (calculated from ordinary 
Levi-Civita connection $\Gamma^{P}_{MN}$). This means that (\ref{10dlcs}) is still not of the 
form (\ref{mixCS}), which we know how to handle in the $\mathcal{E}$-function framework.
Now, it can be shown that the difference is given by \cite{Chemissany:2007he}
\begin{equation} \label{bcscs}
\overline{\Omega}_{MNP} = \Omega_{MNP} + \mathcal{A}_{MNP}
\end{equation}
where
\begin{eqnarray} \label{a6mnp}
 \mathcal{A}_{MNP} &=&
 \frac{1}{4} \partial_M \left( \Gamma^R_{NQ} \overline{H}^{\quad Q}_{RP} \right)
 + \frac{1}{8} \overline{H}^{\quad\; R}_{MQ} \, \nabla_N \overline{H}^{\quad Q}_{RP}
 - \frac{1}{4} R^{\quad\;\; QR}_{MN} \overline{H}_{PQR} \nonumber \\
&& + \frac{1}{24} \overline{H}^{\quad\; R}_{MQ} \overline{H}^{\quad\; S}_{NR}
 \overline{H}^{\quad Q}_{PS} \quad
 \mbox{(antisymmetrized in $M,N,P$).}
\end{eqnarray}
First term in (\ref{a6mnp}) gives vanishing contribution to the action obtained from Lagrangian 
(\ref{10dtl}) due to $dK=0$, and so it can be dropped. It then follows that $\mathcal{A}_{MNP}$ is manifestly diffeomorphism covariant. Using (\ref{bcscs}) in (\ref{10dlcs}) we have
\begin{equation} \label{lcs1p}
\widetilde{\mathcal{L}}_{\rm CS} = \widetilde{\mathcal{L}}'_1 + \widetilde{\mathcal{L}}''_1 \,,
\end{equation}
where
\begin{eqnarray} \label{tl1p}
\widetilde{\mathcal{L}}'_1 &=&
 - \frac{\alpha'}{32\pi^2 (D-3)!\sqrt{-G}} \, \varepsilon^{M_1\cdots M_D} 
 \mathcal{A}_{M_1 M_2 M_3} K_{M_4\cdots M_D} \\
\widetilde{\mathcal{L}}''_1 &=&
 - \frac{\alpha'}{32\pi^2 (D-3)!\sqrt{-G}} \, \varepsilon^{M_1\cdots M_D} 
 \Omega_{M_1 M_2 M_3} K_{M_4\cdots M_D} \;.
\label{tl1pp}
\end{eqnarray}
The term $\widetilde{\mathcal{L}}'_1$ is manifestly diff-covariant, while $\widetilde{\mathcal{L}}''_1$ contains ordinary (Levi-Civita) Chern-Simons term and is obviously of the form (\ref{mixCS}).
This is what we wanted to achieve, so we can now finally pass to calculations.

Our goal is to calculate $\alpha'$-exact solutions and corresponding central charges for various backgrounds in heterotic string theory which have AdS$_3$ and $S^k$, $k=1,2,3$ factors. As we 
shall see a posteriori, all our solutions will satisfy (\ref{modrie0}). It then follows 
(see the discussion below Eq. (\ref{modrie0})) that $\Delta \mathcal{L}_{\rm CS}$ will not 
contribute. We conjecture (based on a limited perturbative knowledge) that 
$\mathcal{L}_{\rm other}$ also should not contribute. It then follows that it is enough to work
with the reduced dual Lagrangian given by
\begin{equation} \label{tlred}
\widetilde{\mathcal{L}}_{\rm red} = \widetilde{\mathcal{L}}_0 + \widetilde{\mathcal{L}}_{\rm CS}
 = \widetilde{\mathcal{L}}_0 + \widetilde{\mathcal{L}}'_1 + \widetilde{\mathcal{L}}''_1 \;.
\end{equation}
Note that $\widetilde{\mathcal{L}}_{\rm red}$ has at most 4-derivative terms (i.e., it is $R^2$-type 
Lagrangian).

We shall use a convention in which $G_D=2$ and $\alpha'=16$.

\subsection{\label{ssec:s3}AdS$\bm{_3 \times S^3}$ backgrounds}

Let us now apply this to $AdS_3 \times S^3$ solutions in heterotic string theory compactified 
on $T^4$. Such backgrounds are expected to describe near-horizon geometries of extremal black
strings in $D=6$ dimensions. The non-vanishing fields here are dilaton $\Phi$, 6-dimensional 
metric $G_{\mu\nu}$, 3-form $\overline{H}_{\mu\nu\rho}$ (treated as auxiliary), and the 2-form 
gauge field $C_{\mu\nu}$ (with 3-form strength $K_{\mu\nu\rho}$). We now use generalized 
version of Sen's $\mathcal{E}$-function formalism presented in Section \ref{sec:senef}, which 
dictates the following form for the non-vanishing fields
\begin{eqnarray}
&& ds^2 = v_A\, ds_A^2 + v_S\, ds_S^2 \,, \qquad e^{-2\Phi} = \frac{u_s}{\pi} \,,
\nonumber \\
&& K = \widetilde{e}\, \epsilon_A + \widetilde{p}\, \epsilon_S \,, \qquad
 \overline{H} = h_A\, \epsilon_A + h_S\, \epsilon_S \,,
\label{nh6d2c}
\end{eqnarray}
where $v_{A,S}$, $u_s$, $h_{A,S}$ are constants, eventually determined from equations of motion 
(as functions of electric field $\widetilde{e}$ and magnetic charge $\widetilde{p}$). These 2-charge 
black string configurations have microscopic interpretation as bound states of some number 
(connected to electric charge of $\overline{H}_{\mu\nu\rho}$, which means magnetic charge of 
$K_{\mu\nu\rho}$) of fundamental strings plus some number (connected to magnetic charge of 
$\overline{H}_{\mu\nu\rho}$, which means electric charge of $K_{\mu\nu\rho}$)) of NS5-branes 
wrapped around the torus $T^4$. 

Using (\ref{tlred}) we can write function $f$, defined in (\ref{fads3}), as
\begin{eqnarray}
f_0 &=& \frac{1}{8} \left[ u_s (v_A v_S)^{3/2} \left( -\frac{3}{v_A} + \frac{3}{v_S}
 + \frac{h_A^2}{4\,v_A^3} - \frac{h_S^2}{4\,v_S^3} \right)
 - h_A\, \widetilde{p} + h_S\, \widetilde{e} \right]
\label{f06d2c} \\
f'_1 &=& \frac{h_A^3\, \widetilde{p}}{4\, v_A^3} + \frac{h_S^3\, \widetilde{e}}{4\, v_S^3}
 - \frac{3\, h_A\, \widetilde{p}}{v_A} - \frac{3\, h_S\, \widetilde{p}}{v_S}
\label{f1p6d2c} \\
f''_1 &=& f_{\rm CS} = \pm 4\,\widetilde{p} \;.
\label{f1pp6d2c}
\end{eqnarray} 
As for derivation and interpretation of (\ref{f1pp6d2c}), consult section 
\ref{ssec:gCS}. $\mathcal{E}$-function, defined in (\ref{efads3}) is now
\begin{equation}
\mathcal{E}(\vec{v},u_s,\vec{h},\widetilde{e};\widetilde{q},\widetilde{p})
 = 6\pi (\widetilde{e}\,\widetilde{q} - f) \;,
\end{equation}
where $\widetilde{q}$ is an electric charge. Extremization of $\mathcal{E}$-function over
$v_{1,2}$, $u_s$, $h_{A,S}$, and $\widetilde{e}$ gives us then conditions equivalent to equations
of motion and central charges, according to (\ref{Eeom})-(\ref{cads3}). The solution is
\begin{equation} \label{sol6d2c}
v_A = v_S = 4(|\widetilde{q}|+4) \,,\quad u_s = \frac{|\widetilde{p}|}{4(|\widetilde{q}|+4)} \,,\quad
\widetilde{e}  = \frac{|\widetilde{q}\,\widetilde{p}|}{\widetilde{q}} \,,\quad  
h_A = 8 \frac{|\widetilde{p}|}{\widetilde{p}} \left( |\widetilde{q}| + 4 \right) \,,\quad
h_S = 8 \frac{|\widetilde{q}|}{\widetilde{q}} \left( |\widetilde{q}| + 4 \right)
\end{equation}
which is valid for all $\widetilde{q}$ and $\widetilde{p}\ne0$.
In the special cases when $\widetilde{q}=0$ it is understood that $|0|/0=\pm1$, meaning that there 
are two solutions for fixed choice of $\widetilde{p}$.

For central charges we obtain
\begin{equation} \label{c6d2c}
c \equiv \frac{1}{2} (c_L + c_R) = 6\pi\,|\widetilde{p}| \left( |\widetilde{q}| + 8 \right) \;,\qquad
c_L - c_R = 48\pi\,\widetilde{p}
\end{equation}

We still have to connect "canonical" charges $\widetilde{q}$, $\widetilde{p}$ with integer-valued
charges of microscopic configuration (consisting of fundamental strings and NS5-branes). Normally,
one does this by referring to the (if known) lowest-order relations. In the present case, these are 
well-known, and, in our conventions, are given by
\begin{equation} \label{pw6d2c}
w = 4\pi\,\widetilde{p} \;, 
\end{equation}
where microscopic charge $w$ is the number of fundamental strings, and
\begin{equation} \label{qm6d2c}
m = \frac{\widetilde{q}}{4} \;, 
\end{equation}
where $m$ is the NS5-brane charge. Using (\ref{pw6d2c}) and (\ref{qm6d2c}) in (\ref{sol6d2c}) 
and (\ref{c6d2c}) we obtain for solution
\begin{eqnarray} \label{solm6d2c}
&& v_A = v_S = 16(|m|+1) \,,\quad u_s = \frac{1}{64\pi} \frac{|w|}{(|m|+1)} \,,\quad
 \widetilde{e}  = \frac{|wm|}{4\pi m} \,,
\nonumber \\  
&& h_A = 32 \frac{w}{|w|} (|m|+1) \,,\quad h_S = 32 \frac{m}{|m|} (|m|+1) \,,
\end{eqnarray}
and for central charges
\begin{equation} \label{cm6d2c}
c \equiv \frac{1}{2} (c_L + c_R) = 6\,|w| (|m| + 2) \;,\qquad  c_L - c_R = 12\,w
\end{equation}

In the theories with Chern-Simons terms things can get more complicated, as it is known that such
terms may introduce shifts between charges defined near the horizon and charges defined 
in asymptotic infinity (which is standard definition for charges). This effect was previously observed
and analyzed in black hole setup in \cite{Castro:2007ci,deWit:2009de}. Now, one way which avoid
such issues to express all charges as magnetic charges of some gauge form with closed field 
strength. For $\widetilde{p}$ this is already done, as it is the magnetic charge of a 
closed 3-form strength $K$, so we do not expect corrections to (\ref{pw6d2c}). But,
$\widetilde{q}$ is electric charge of $K$, so some additional work is necessary. 

In \cite{Prester:2008iu} it was proposed that one should use 3-form strength $\overline{H}$. From
(\ref{hbcs}) follows
\begin{equation} \label{dHbar}
d\overline{H} = \frac{3}{8} \alpha' \mbox{tr} (\overline{R} \wedge \overline{R})
\end{equation}
and because all our solutions satisfy condition (\ref{modrie0}) we have $d\overline{H}=0$. 
Integer-valued magnetic charge carried by $\overline{H}$ is given by\footnote{The factor of 
$64\pi^2$, which appears in our conventions, is in fact $2\alpha'\Omega_3$ ($\Omega_3$ is the 
volume of a unit 3-sphere $S^3$).}
\begin{equation} \label{Hbarmc}
N = \frac{1}{64\pi^2} \oint_{S^3} \overline{H} = \frac{h_S}{32} = \frac{m}{|m|} (|m| + 1) \,.
\end{equation}
We obtain a shift in the definition of charge. In a new definition of charge (\ref{Hbarmc}) doubling 
of solutions for $m=0$ becomes natural -- from (\ref{Hbarmc}) follows that $m=0$ simply 
corresponds to two values of $N$, $N=\pm1$. Using (\ref{Hbarmc}) in (\ref{solm6d2c}) and 
(\ref{cm6d2c}) we obtain for solution
\begin{equation} \label{solf6d2c}
v_A = v_S = 16\,|N| \;,\quad u_s = \frac{1}{64\pi} \left| \frac{w}{N} \right| \;,\quad
\widetilde{e} = \frac{|w N|}{4\pi\,N} \;,\quad  h_A = 32 \frac{|w\, N|}{w} \;,\quad
h_S = 32\, N \;,
\end{equation}
and for central charges
\begin{equation} \label{cf6d2c}
c \equiv \frac{1}{2} (c_L + c_R) = 6\,|w| (|N| + 1) \;,\qquad
c_L - c_R = 12\,w \;.
\end{equation}
The result for central charges $c_{R,L}$ is the same as the one obtained in \cite{Kutasov:1998zh}, 
if we identify $N$ with quantum number $k$ from \cite{Kutasov:1998zh}. 

However, there is a problem with the above definition of charge, which is visible for more 
complicated geometries.\footnote{We thank Ashoke Sen for thorough explanations on this issue 
and for suggestion that 3-form $H$, instead of $\overline{H}$, should be used to define the proper
asymptotic charge.} As a specific example, relevant for us here, let us consider a full black string background for which 
(\ref{solm6d2c}) gives a near-horizon description. Away from the horizon we do not expect 
(\ref{modrie0}) to be valid, and so by (\ref{dHbar}) there is no reason to expect $d\overline{H}=0$. 
Without this property, there is no guarantee that "near-horizon charge" (\ref{dHbar}), which is 
calculated by taking $S^3$ to be in the near-horizon region where (\ref{solm6d2c}) is 
(approximately) valid, is going to be equal to the standardly defined (asymptotic) charge, which is 
obtained by taking sphere $S^3$ in (\ref{Hbarmc}) to be in an asymptotic infinity.

A simple solution for this problem is that instead of $\overline{H}$, defined in (\ref{hbcs}), we use
a 3-form $H$ defined by 
\begin{equation} \label{hcs}
H_{MNP} = \partial_M B_{NP} + \partial_N B_{PM} + \partial_P B_{MN}
  - 3 \alpha' \Omega_{MNP} \;,
\end{equation}
which is in fact "standard" definition for 3-form strength of Kalb-Ramond field in heterotic string 
theory. The important difference is that $dH$ is given by 
\begin{equation} \label{dH}
dH = \frac{3}{8} \alpha' \mbox{tr} (R \wedge R) \,,
\end{equation}
and the right hand side is now topological density which is giving vanishing contribution to the 
difference between asymptotic and near-horizon charges.
From (\ref{bcscs}) follows that $H$ and $\overline{H}$ are connected by
\begin{equation} \label{hhbar}
H_{MNP} = \overline{H}_{MNP} + 3 \alpha' \mathcal{A}_{MNP} \;.
\end{equation}
From (\ref{hhbar}) and (\ref{a6mnp}) we easily obtain a corresponding magnetic 
charge
\begin{equation} \label{Hmc}
Q_5 = \frac{1}{64\pi^2} \oint_{S^3} H = m \,,
\end{equation}
which, though it is near-horizon evaluated, is also equal to the standard (asymptotic) charge. We 
emphasize that $H$ and $\overline{H}$ give the same result for charge calculated in 
infinity\footnote{The "dangerous" first term in (\ref{a6mnp}), which is not manifestly diff-covariant, 
is a total derivative and gives vanishing contribution to the integral in (\ref{Hmc}). The rest of
the difference between $H$ and $\overline{H}$ is such that its contribution to the charge 
calculated in asymptotic infinity vanishes.}, and the difference in near-horizon charge 
accumulates in the "intermediate region" when one passes from infinity towards horizon. 

All in all, we finally obtained that our original definition of electric charge $\widetilde{q}=4m$, which 
depends on the particular treatment of mixed Chern-Simons term, is correctly describing the
magnetic flux, and so $Q_5 = m$ is expected to be identified with the number of NS5-branes.

Comments on our AdS$_3 \times S^3 \times T^4$ solution:
\begin{enumerate}
\item
Solution (\ref{solf6d2c}) is supersymmetric for all values of charges.
\item
It is easy to show that solution (\ref{solf6d2c}) satisfies the property (\ref{modrie0}), which means 
that we would obtain the same results if we started with more complicated supersymmetric action 
(\ref{susyl}). Moreover, as argued before (and in more detail in section \ref{sec:concl}), we suggest
that (\ref{modrie0}) also makes $\mathcal{L}_{\rm other}$ is also irrelevant for our results, which 
means that we have obtained $\alpha'$-exact solutions and central charges of the full tree-level 
heterotic effective action (\ref{10dl}).
\item
Though solution (\ref{solf6d2c}) is purely mathematically regular for all $w\ne0$, from string theory
perspective it is meaningful only for $|w/m|\gg1$ as in this case quantum corrections are expected
to be small (effective string coupling $g_s$ satisfies $g_s^2 \sim \exp(2\Phi) \sim |N/w| \ll1$)
and so our purely classical analysis is dominating.
% \item
% Solution is singular for $N=0$ (small black string). This is not surprising because we see from
% (\ref{solf6d2c}) that $\alpha'$-expansion is effectively $1/N$ expansion, and so it is not well 
% definedwhen $N=0$. In this case microscopic configuration is consisting just of fundamental 
% string, for which we normally do not expect to be described well by low-energy effective action.
\item
Results for central charges $c_{L,R}$ agree with microscopic calculations relying on 
AdS$_3$/CFT$_2$ arguments \cite{Kutasov:1998zh}, when we identify our number $N$ 
(near-horizon charge of 3-form strength $\overline{H}$) with their $k$, which is a level of the affine 
world-sheet symmetry algebra $\widehat{SL(2)}$ in the supersymmetric (right-moving) sector. Also, 
using $N$ (instead of $m$) we obtain solutions in $\alpha'$-uncorrected form.
\item
In view of possible AdS/CFT correspondence, our results for central charges $c_{L,R}$ agree 
(through the Cardy formula)) with results for entropies of 5-dimensional 3-charge extremal black 
hole entropies calculated in \cite{Prester:2008iu}, while comparison with calculation which uses 
$R^2$-type 5-dimensional supersymmetric action shows agreement just for the cases which
correspond to BPS black holes (e.g., for $w,m > 0$ it agrees with $c_L$, but not with $c_R$) 
\cite{Cvitan:2007en} (for BPS case it exist also a microscopic calculation for the entropy 
\cite{Castro:2008ys}). 
\end{enumerate}

\subsection{\label{ssec:s2}AdS$\bm{_3 \times S^2}$ backgrounds}

The next example are $AdS_3 \times S^2$ solutions, which should describe near-horizon 
geometries of extremal black strings in $D=5$ dimensions. We start from heterotic string theory compactified on $S^1 \times T^4$, taking for the charges coming from Kaluza-Klein fields of 
$S^1$ reduction to be non-vanishing. The details of this particular Kaluza-Klein reduction are 
reviewed in \cite{PDPlosinj}. In our notation coordinate radius of $S^1$ is $\sqrt{\alpha'}=4$. 
The non-vanishing fields are: string metric $G_{\mu\nu}$, dilaton $\Phi$, modulus 
$T=(\widehat{G}_{55})^{1/2}$, two Kaluza-Klein gauge fields $A_{(i)\mu}$ ($1\le i\le 2$), coming 
from $G^{(6)}_{MN}$ and 2-form potential $C^{(6)}_{MN}$, the 2-form potential $C_{\mu\nu}$ with 
the strength $K_{\mu\nu\rho}$, one Kaluza-Klein auxiliary two form $\overline{D}_{\mu\nu}$ coming 
from $\overline{H}^{(6)}_{MNP}$, and auxiliary 3-form $\overline{H}_{\mu\nu\rho}$.\footnote{Greek 
indices are 5-dimensional, i.e., $0\le\mu,\nu,\ldots\le4$, while capital latin indices are 
6-dimensional, i.e., $0\le M,N,\ldots\le5$. Coordinate on $S^1$ is denoted $x^5$, with 
$0 \le x^5 < 8\pi$.}  
$\mathcal{E}$-function formalism then dictates the following form for the non-vanishing fields
\begin{eqnarray}
&& ds^2 = G_{\mu\nu} dx^\mu dx^\nu = v_A\, ds_A^2 + v_S\, ds_S^2 \,, \qquad
 e^{-2\Phi} = u_s \,,\qquad T = u_t \,,
\nonumber \\
&& K = \frac{\widetilde{e}}{8}\, \epsilon_A \,, \qquad F_1 = \widetilde{p}_1 \, \epsilon_S \,,\qquad
 F_2 = \frac{\widetilde{p}_2}{16} \, \epsilon_S \,,\qquad
 \overline{H} = h_A\, \epsilon_A \,,\qquad \overline{D} = - \frac{d_S}{2}\, \epsilon_S \,,
\label{nh5d3c}
\end{eqnarray}
where now $\epsilon_S$ is a volume-form of unit $S^2$ sphere, $v_{A,S}$, $u_s$, $u_t$, $h_A$ 
and $d_S$ are constants, eventually determined from equations of motion (as functions of electric 
field $\widetilde{e}$ and magnetic charges $\widetilde{p}_{1,2}$). These 3-charge black string configurations have microscopic interpretation as bound states of some number (connected to 
electric charge of $\overline{H}_{\mu\nu\rho}$, which means magnetic charge of $F_{(2)\mu\nu}$) 
of fundamental strings, some number (connected to magnetic charge of $\overline{D}_{\mu\nu}$, 
which means electric charge of $F_{(1)\mu\nu}$) of NS5-branes wrapped around torus $T^4$, and 
some number (connected to magnetic charge of $\overline{D}_{\mu\nu}$, which means electric 
charge of $K_{\mu\nu\rho}$) of Kaluza-Klein monopoles with "nut" on $S^1$.

As originally proposed in \cite{Sahoo:2006pm}, the most efficient way to calculate the 
$\mathcal{E}$-function is to lift 5-dimensional background (\ref{nh5d3c}) back to 6-dimensions
(by using KK reduction relations backwards) and than to perform calculation of $f$ function in 
6-dimensions, where the action has much simpler form (presented in section \ref{ssec:10Dha}). 
Details of this KK reduction are reviewed in \cite{PDPlosinj}. By using them the background 
(\ref{nh5d3c}) in 6-dimensional language becomes
\begin{eqnarray}
&& ds^2 = G^{(6)}_{MN} dx^M dx^N
 = v_A\, ds_A^2 + v_S\, ds_S^2 + u_t^2 \left( dx^5 - 2\,\widetilde{p}_1 \cos\theta\, d\phi \right)^2
 \,, \qquad e^{-2\Phi^{(6)}} = \frac{u_s}{8\pi\, u_t} \,,
\nonumber \\
&& K^{(6)}_{012} = \frac{\widetilde{e}}{8}\,, \qquad
 K^{(6)}_{\theta\phi5} = - \frac{\widetilde{p}_2}{8} \sin\theta \,,\qquad
 \overline{H}^{(6)}_{012} = h_A \,,\qquad \overline{H}^{(6)}_{\theta\phi5} = d_S \sin\theta \,.
\label{nh6d3c}
\end{eqnarray}

Again, using (\ref{tlred}) we can write function $f$, defined in (\ref{fads3}), as
\begin{equation} \label{f5d3c}
f(\vec{v},\vec{u},\vec{h};\widetilde{e},\vec{\widetilde{p}})
 = \oint_{S^2 \times S^1} \sqrt{-G^{(6)}} \, \widetilde{\mathcal{L}}_{\rm red}^{(6)}
 = f_0 + f'_1 + f''_1 \,,
\end{equation}
Using (\ref{nh6d3c}) we obtain
\begin{eqnarray}
f_0 &=& \frac{1}{4} \left[ u_s\, v_A^{3/2}\, v_S \left( -\frac{3}{v_A} + \frac{1}{v_S}
 - \frac{u_t^2\,\widetilde{p}_1^2}{v_S^2} - \frac{d_S^2}{4\, u_t^2\, v_S^2}
 + \frac{h_A^2}{4\, v_A^3} \right)
 + h_A\, \widetilde{p}_2 + u_t^2\, d_S\, \widetilde{e} \right]
\label{f05d3c} \\
f'_1 &=& \frac{6\, h_A\, \widetilde{p}_2}{v_A} - \frac{h_A^3\, \widetilde{p}_2}{2\, v_A^3}
 + \frac{d_S^3\, \widetilde{e}}{2\, u_t^2\, v_S^2} + \frac{2\, u_t^2\, d_S\, \widetilde{p}_1^2}{v_S^2}
 - \frac{2\, d_S\, \widetilde{e}}{v_S}
\label{f1p5d3c} \\
f''_1 &=& \pm 8\,\widetilde{p}_2 + 4\, \widetilde{e}
 \left( \frac{u_t^2\, \widetilde{p}_1}{v_S} - 2 \frac{u_t^4\, \widetilde{p}_1^3}{v_S^2} \right) \;.
\label{f1pp5d3c}
\end{eqnarray}
where for practical purposes we passed to variables $h_{1,2}$, defined by
\begin{equation}
h_1 \equiv - \frac{u_s\, v_2}{2\, v_1^{3/2}} h_A \,, \qquad
h_2 \equiv \frac{u_s\, v_1^{3/2}}{2\, u_t^2\, v_2} d_S \,,
\end{equation}
instead of $h_A$ and $d_S$. $\mathcal{E}$-function is given by
\begin{equation}
\mathcal{E}(\vec{v},\vec{u},\vec{h},\widetilde{e};\widetilde{q},\vec{\widetilde{p}})
 = 6\pi \left( \widetilde{e}\, \widetilde{q} - f \right) \,,
\end{equation}
where $\widetilde{q}$ is electric charge conjugated to $\widetilde{e}$. By extremizing  
$\mathcal{E}$-function over $v_{A,S}$, $u_{t,s}$, $h_A$, $d_S,$ and $\widetilde{e}$ we obtain 
the solutions. Before writing them down, let us make connection between canonical charges 
$\widetilde{q}$, $\widetilde{p}_{1,2}$ and integer-valued microscopic charges. As all U(1) gauge-field
strengths are closed, we can safely use lowest-order relations which in our conventions read
\begin{equation} \label{chcm5}
\widetilde{q} = - \frac{W'}{2} \;,\qquad \widetilde{p}_1 = N' \;,\qquad 
\widetilde{p}_2 = \frac{w}{8\pi} \;.
\end{equation}
In microscopic interpretation (of black string) $w$ is the number of fundamental strings, $N'$ is the 
Kaluza-Klein monopole charge, and $W'$ is the NS5-brane charge.

Supersymmetric solutions, characterized by $N'W'\ge0$, are given by
\begin{eqnarray}
&& v_A = 4\,v_S = 16(N'W'+2) \;,\qquad u_s = \frac{1}{8\pi} \frac{|w|}{\sqrt{N'W'+2}} \;,\qquad
 u_t = \sqrt{\frac{W'}{N'} \left( 1 + \frac{2}{N'W'} \right)} \;,
\nonumber \\
&& \widetilde{e} = - \frac{|w N'W'|}{\pi\, W'} \;,\qquad 
 h_A = 32 \frac{w}{|w|} (N'W'+2) \;,\qquad
 d_S = 4\, W' \left( 1 + \frac{2}{N'W'} \right) \;.
\label{solf5d3c}
\end{eqnarray}
Central charges in BPS case are
\begin{equation} \label{cf5d3c}
c \equiv \frac{1}{2} (c_L + c_R) = 6\,|w| (N'W' + 3) \;,\qquad
c_L - c_R = 12\,w \;.
\end{equation}
For $w>0$ (\ref{cf5d3c}) gives
\begin{equation} \label{cs5d3c}
c_L = 6\,|w| (N'W' + 4) \;,\qquad c_R = 6\,|w| (N'W' + 2) \;,
\end{equation}
while for $w<0$ one just has to exchange $c_L \leftrightarrow c_R$.

Non-supersymmetric solutions, characterized by $N'W'<0$, are given by
\begin{eqnarray}
&& v_A = 4\,v_S = 16|N'W'| \;,\qquad u_s = \frac{1}{8\pi} \frac{|w|}{\sqrt{|N'W'|}} \;,\qquad
 u_t = \sqrt{\left| \frac{W'}{N'} \right|} \;,
\nonumber \\
&& \widetilde{e} = - \frac{|w N'W'|}{\pi\, W'} \;,\qquad 
 h_A = 32 \frac{|w N'W'|}{w} \;,\qquad
 d_S = 4\, W' \;.
\label{solfn5d3c}
\end{eqnarray}
Central charges in non-BPS case are
\begin{equation} \label{cfn5d3c}
c \equiv \frac{1}{2} (c_L + c_R) = 6\,|w| (|N'W'| + 1) \;,\qquad
c_L - c_R = 12\,w \;.
\end{equation}
For $w>0$ (\ref{cfn5d3c}) gives
\begin{equation} \label{csn5d3c}
c_L = 6\,|w| (|N'W'| + 2) \;,\qquad c_R = 6\,|w N'W'| \;,
\end{equation}
while for $w<0$ one again just has to exchange $c_L \leftrightarrow c_R$.

It is interesting to find charges calculated from fluxes of 10-dimensional 3-forms $\overline{H}$ and 
$H$. In the case of $\overline{H}$, which is defined in (\ref{hbcs}), the corresponding charge 
$\overline{W}$ is calculated from  
\begin{equation} \label{Wbdef}
\overline{W} \equiv \frac{1}{128\,\pi^2} \oint_{S^2 \times S^1} \overline{H} = \frac{h_S}{4} 
\end{equation}
In BPS case, using (\ref{solf5d3c}), we obtain
\begin{equation} \label{WbWpb}
\overline{W} =  W' + \frac{2}{N'} 
\end{equation}
while in the non-BPS case we obtain a simple uncorrected relation
\begin{equation} \label{WbWpn}
\overline{W} = W'
\end{equation}
Using $\overline{W}$ instead of $W'$ puts all solutions (BPS and non-BPS) in $\alpha'$-uncorrected 
form, and central charges $c_{R,L}$ in the form (\ref{cfn5d3c}).

However, we argued in section \ref{ssec:s3} that to obtain proper asymptotic charge, instead of 
$\overline{H}$ we should use 3-form strength $H$ defined in (\ref{hcs}). The corresponding flux
quantum number is now
\begin{equation} \label{Q5def}
Q_5 \equiv \frac{1}{128\,\pi^2} \oint_{S^2 \times S^1} H \, 
\end{equation}
which again can be obtained by using (\ref{hhbar}) and (\ref{a6mnp}). The result is
\begin{equation} \label{Q5Wp}
Q_5 = W' + \frac{1}{N'} 
\end{equation}
In the case $N'=1$ we obtain $Q_5 = W'+1$. Now, it is known 
\cite{Bershadsky:1995qy,Sen:2007qy} that presence of one Kaluza-Klein monopole adds 
$(-1)$-unit to NS5-brane charge, so in this case we can again identify of $Q_5$ with the number of 
NS5-branes. 

Comments on our AdS$_3 \times S^2 \times S^1 \times T^4$ solutions (\ref{solf5d3c}) and 
(\ref{solfn5d3c}):
\begin{enumerate}
\item
It is easy to show that both solutions satisfy property (\ref{modrie0}). Consequences of this are
the same as in section \ref{ssec:s3}.
\item
Though our solutions are regular for all $|w|\ne0$, from string theory perspective they are 
meaningful only for $|w|/\sqrt{|N'W'|} \gg 1$ as in this case quantum corrections are expected to be 
small (effective string coupling $g_s^2 \sim \exp(2\Phi) = 1/u_s \ll1$) and so our purely classical 
analysis is indeed dominant.
\item
It is obvious that $\alpha'$-expansion is here effectively $1/|N'W'|$ expansion. So, one would expect problems for $N'=0$ and/or $W'=0$. However, we see that (\ref{solf5d3c}) is completely regular for $W'=0$, $N'\ne0$, though it is singular when $N'=0$. Now, this is a bit strange
because heterotic theory has a particular T-duality on $N' \longleftrightarrow W'$ (in which one 
expects that $T \to 1/T$ and $F_{(1)\mu\nu} \longleftrightarrow D_{\mu\nu}$), which now 
appears to be broken. This is of course not the case, and the resolution is that in non-trivial 
$S^1$ compactifications higher derivative corrections can change the relations between canonical 
fields (in our case $T$, $F_{(1)\mu\nu}$ and $D_{\mu\nu}$) and proper string moduli (in our case 
$S^1$ radius $R$ and fluxes $\mathcal{F}_{(1)\mu}$, $\mathcal{D}_{\mu\nu}$), and one needs to 
find appropriate field redefinitions before making identifications (see, e.g., \cite{Sen:2004dp}). 
\item
Our results for central charges agree with microscopic calculations relying on AdS$_3$/CFT$_2$ arguments \cite{Kutasov:1998zh,Kraus:2005vz}. Again, charge $\overline{W}$ obtained from 3-form 
$\overline{H}$ is connected with the total level $k$ of the world-sheet affine algebra 
$\widehat{SL(2)}$ in the supersymmetric sector, and the relation is now $k=N'\overline{W}$. Also,
use of $\overline{W}$ puts solutions in the $\alpha'$-uncorrected form. 
\item
Agreement with microscopic calculations is now also true for $N'=W'=0$, where solutions describe 
near-horizon geometry of \emph{small} black (fundamental) string. We 
mentioned above that in this regime low energy/curvature effective action is not expected to be 
well defined (as $\alpha'$-expansion is not well defined).\footnote{Though there is a proposal 
that near-horizon properties of small black strings could be effectively described by a simple 
Lovelock-type action \cite{Prester:2005qs}. Analysis from \cite{Cai:2007cz} is also giving some 
wind for this proposal. The extension of small black hole solutions to the whole space-time in the 
case of Gauss-Bonnet-type action was analyzed in \cite{Chen:2009rv}.} This agreement is in 
contrast with the case of 6-dimensional black string (analyzed in previous section) where putting 
$m=0$ gives wrong results for central charges.  
% On the other hand, in 5-dimensional case small black string limit ($N'=W'=0$) of (\ref{solf5d3c}) is 
% \emph{regular} for $v_{A,S}$, $u_s$, $h_A$ and $\widetilde{e}$, and singular only for $u_t$ and 
% $d_S$. We believe that these singularities are not intrinsic and can be removed by the clever field 
% redefinitions, that would at the same time make $u_t$ to correspond to the true modulus of $S^1$ 
% (which should manifestly respect T-duality mentioned above). We shall address this issue in the 
% separate publication.
\item
The same result for $c$ in BPS case (\ref{cf5d3c}) was also obtained from 5-dimensional 
$R^2$-type action obtained by off-shell supersymmetrization of mixed Chern-Simons term in 
\cite{Castro:2007sd}. However, this action produces wrong result for $c$ in non-BPS case, 
deviating from (\ref{cfn5d3c}) already at $\alpha'^1$-order.\footnote{This can be easily shown by 
simple extension of the method for constructing non-BPS solutions from \cite{Cvitan:2007en} to 
AdS$_3 \times S^2$ geometries.} 
This shows that this $R^2$-supersymmetric action is incomplete already at 4-derivative order, a
fact already noted in \cite{Cvitan:2007en,Prester:2008iu}.
\item
Using AdS/CFT correspondence, our results for central charges $c_{L,R}$ agree 
(through the Cardy formula)) with results for entropies of 4-dimensional 4-charge extremal black 
hole entropies calculated in \cite{Prester:2008iu}. Comparison with calculation which uses 
$R^2$-type 4-dimensional supersymmetric action shows agreement just for the cases which
correspond to BPS black holes \cite{Cvitan:2007en}.
\item
For $N'=1$, 3-charge AdS$_3 \times S^3 \times T^4$ solutions become equal to 2-charge 
AdS$_3 \times S^2 \times S^1 \times T^4$ solutions when one identifies the corresponding $Q_5$
charges, i.e., for $m=W'+1$. The explanation of this comes from understanding of these backgrounds 
as near-horizon geometries of black strings. Microscopic explanation of this "charge-shift" was given 
in \cite{Bershadsky:1995qy}, while macroscopic explanation, in the framework of $R^2$-type 
supersymmetric effective action, was given in \cite{Castro:2007ci}. In our analysis (and also in
\cite{Prester:2008iu}), charge-shift appears because Lorentz Cher-Simons term is
evaluated on different topologies ($S^3$ in one case, and $S^2\times S^1$ in the 
other).\footnote{Note that geometries are locally isomorphic, but not globally.}  
\end{enumerate}

\subsection{\label{ssec:s3s3}AdS$\bm{_3 \times S^3 \times S^3}$ backgrounds}

Our final example are AdS$_3 \times S_+^3 \times S_-^3$ solutions of the heterotic string theory
compactified on $S^1$. Compactification on $S^1$ is trivial (corresponding KK charges are
all zero), and $\pm$ subscript on $S^3$ is put just to separate two 3-spheres. Contrary to previous
two examples, these backgrounds do not have direct interpretation as near-horizon geometries of
some black objects (and so, e.g., are not listed in \cite{Duff:2008pa}). Calculations here are 
similar to those from section \ref{ssec:s3}, with the difference that now we have two 3-spheres and 
$K$ is a 6-form (because effective space-time is 9-dimensional).

Now, $\mathcal{E}$-function formalism forces the following form for the non-vanishing fields
\begin{eqnarray}
&& ds^2 = v_A\, ds_A^2 + v_+\, ds_+^2 + v_-\, ds_-^2 \,, \qquad S = u_s \,,
\label{nh9d3c} \\
&& K = \widetilde{e}_-\, \epsilon_A \wedge \epsilon_+
 + \widetilde{e}_+\, \epsilon_A \wedge \epsilon_-
 +  \widetilde{p}\, \epsilon_+ \wedge \epsilon_- \,,\qquad
  H = h_A\, \epsilon_A + h_+\, \epsilon_+ + h_-\, \epsilon_- \,,
\nonumber
\end{eqnarray}
Function $f$ is now
\begin{equation} \label{f9d3c}
f(\vec{v},u_s,\vec{h};\widetilde{e},\widetilde{p})
 = \oint_{S_+^3 \times S_-^3} \sqrt{-G} \, \widetilde{\mathcal{L}}_{\rm red}
 = f_0 + f'_1 + f''_1 \,,
\end{equation}
Using (\ref{nh9d3c}) we obtain
\begin{eqnarray}
f_0 &=& \frac{\pi^3}{4} \left[ u_s (v_A v_+ v_-)^{3/2} \left(
 \frac{3}{v_+} + \frac{3}{v_-} - \frac{3}{v_A} 
 + \frac{h_A^2}{4\, v_A^3} -  \frac{h_+^2}{4\, v_+^3} -  \frac{h_-^2}{4\, v_-^3} \right)
 - h_A\, \widetilde{p} + h_+\, \widetilde{e}_+ + h_-\, \widetilde{e}_-  \right] \quad
\label{f09d3c} \\
f'_1 &=& \frac{\pi^3}{2} \left( \frac{h_A^3\, \widetilde{p}}{v_A^3}
 + \frac{h_+^3\, \widetilde{e}_+}{v_+^3} + \frac{h_-^3\, \widetilde{e}_-}{v_-^3} \right)
 - 6\pi^3 \left(
 \frac{h_A\, \widetilde{p}}{v_A} + \frac{h_+\, \widetilde{e}_+}{v_+} + \frac{h_-\, \widetilde{e}_-}{v_-}
 \right)
\label{f1p9d3c} \\
f''_1 &=& f_{\rm CS} = \pm 48\pi^4\,\widetilde{p} \;.
\label{f1pp9d3c}
\end{eqnarray} 
$\mathcal{E}$-function is now defined by
\begin{equation}
\mathcal{E}(\vec{v},u_S,\vec{h},\vec{\widetilde{e}};\vec{\widetilde{q}},\widetilde{p})
 = 6\pi \left( \widetilde{e}_+\, \widetilde{q}_+ + \widetilde{e}_-\, \widetilde{q}_- - f \right) \,.
\end{equation}
Extremization of  $\mathcal{E}$-function over $v_{A,\pm}$, $u_s$, $h_{A,\pm}$ and 
$\widetilde{e}_\pm$ gives us the following solution
\begin{eqnarray}
&& v_\pm = \frac{2}{\pi^3} |\widetilde{q}_\pm| + 16 \;,\qquad
 v_A = \frac{v_+\, v_-}{v_+ + v_-} \;,\qquad
 u_s = \frac{v_A^{1/2}\, |\widetilde{p}|}{(v_+\, v_-)^{3/2}} \;,\qquad
\nonumber \\
&& \widetilde{e}_\pm = \frac{|\widetilde{q}_\pm \, \widetilde{p}|}{\widetilde{q}_\pm}
 \left( \frac{v_A}{v_\pm} \right)^{\!2} \;,\qquad 
 h_A = 2\, v_A \frac{\widetilde{p}}{|\widetilde{p}|}\;,\qquad
 h_\pm = 32\, \frac{q_\pm}{|q_\pm|} \left( \frac{|q_\pm|}{8\pi^3} + 1 \right)
\label{sol9d3c}
\end{eqnarray}
From the expression for $h_\pm$ in (\ref{sol9d3c}) we can read that integer-valued fluxes 
$N_5^\pm$, corresponding to 3-form $\overline{H}$, through 3-spheres $S_\pm^3$ are given by
\begin{equation} \label{fluxQpm}
N_5^\pm = \frac{q_\pm}{|q_\pm|} \left( \frac{|q_\pm|}{8\pi^3} + 1 \right) \;.
\end{equation}
The remaining integer-valued charge $Q_1$ is given by the well-known lowest-order relation 
(see, e.g., \cite{Gukov:2004ym})
\begin{equation} \label{Q1p}
Q_1 = 8\pi^4\, \widetilde{p}
\end{equation}
Using (\ref{fluxQpm}) and (\ref{Q1p}) in (\ref{sol9d3c}) we finally obtain that 
AdS$_3 \times S^3 \times S^3$ solution is given by
\begin{eqnarray}
&& v_\pm = 16\, |N_5^\pm| \;,\qquad
 v_A = 16 \frac{|N_5^+\, N_5^-|}{|N_5^+| + |N_5^-|} \;,\qquad
 u_s = \frac{1}{2(8\pi)^4} \left| \frac{Q_1}{N_5^+\, N_5^-} \right| \left( |N_5^+| + |N_5^-| \right)^{-1/2}
  \;,\qquad
\nonumber \\
&& \widetilde{e}_\pm = \frac{1}{8\pi^4} \frac{|N_5^\pm \, Q_1|}{N_5^\pm}
 \left( \frac{v_A}{v_\pm} \right)^{\!2} \;,\qquad 
 h_A = 2\, v_A \frac{Q_1}{|Q_1|}\;,\qquad
 h_\pm = 32\, N_5^\pm
\label{solf9d3c}
\end{eqnarray}
We see that by using 3-form $\overline{H}$ to define magnetic charges, solutions again have 
$\alpha'$-uncorrected form, and the sole effect of $\alpha'$-corrections are charge-shifts 
(\ref{fluxQpm}).

Finally, the central charges are given by
\begin{equation} \label{cfn9d3c}
c \equiv \frac{1}{2} (c_L + c_R)
 = 6\, |Q_1| \left( \frac{|N_5^+\, N_5^-|}{|N_5^+| + |N_5^-|} + 1 \right) \;,\qquad
c_L - c_R = 12\, Q_1 \;.
\end{equation}
For $Q_1>0$ (\ref{cfn9d3c}) reads
\begin{equation} \label{csn9d3c}
c_L = 6\, |Q_1| \left( \frac{|N_5^+\, N_5^-|}{|N_5^+| + |N_5^-|} + 2 \right) \;,\qquad
c_R = 6\, |Q_1|  \frac{|N_5^+\, N_5^-|}{|N_5^+| + |N_5^-|} \;,
\end{equation}
while for $Q_1<0$ one just has to exchange $c_L \leftrightarrow c_R$.

If we instead use 3-form $H$ to calculate charges $Q_5^\pm$, we obtain
\begin{equation}
|Q_5^\pm| = |N_5^\pm| -1
\end{equation}

Comments on our AdS$_3 \times S^3 \times S^3 \times S^1$ results:
\begin{enumerate}
\item
It is easy to show that solution (\ref{sol6d2c}) satisfies property (\ref{modrie0}). Consequences of 
this are the same as in sections \ref{ssec:s3} and \ref{ssec:s2}.
\item
Though solution (\ref{sol6d2c}) is regular for all $Q_1\ne0$, from string 
theory perspective it is meaningful only for $|Q_1| \gg |N_5^+\, N_5^-| ( |N_5^+| + |N_5^-| )^{1/2}$, because then string coupling satisfies $g_s^2 \sim \exp(2\Phi)\ll1$ and our purely classical analysis
is valid.
\item
Solution is singular for vanishing $N_5^+$ or $N_5^-$. Again, this is not surprising because we 
see from our solution that $\alpha'$-expansion is effectively expansion in $1/N_5^+$, 
$1/N_5^-$,\footnote{More precisely, $|N_5^+|^n |N_5^-|^m$ terms will appear at $\alpha'^{n+m}$ 
order.} and so it is not well defined when any of the charges vanish.
\item
As far as we know, our results for central charges are new. In particular, we are not aware of any
$\alpha'$-exact microscopic calculation of central charges in this case. In fact, even the 
microscopic configuration of strings/branes which should lead to such backgrounds is not known. 
Also, a holographic (CFT$_2$) dual is still not known\footnote{See \cite{Gukov:2004ym} for 
thorough analysis of this issue in type-II string theories.}, contrary to previous examples analyzed 
in sections \ref{ssec:s3} and \ref{ssec:s2}.
\item
Solution is supersymmetric for all values of charges.
\end{enumerate}

\subsection{\label{ssec:typeII}Solutions in type-II superstring theories}

All geometries we considered in the paper also appear in NS-NS sector of type-II string theories. 
We now show that from our analysis directly follows that such type-II solutions will all be 
$\alpha'$-\emph{uncorrected}. The reason is that in type-II theories there are no classical Lorentz 
Chern-Simons terms (and in particular they are not present in (\ref{hbcs})), and the truncated 
tree-level effective action is given by
\begin{equation}
\mathcal{L}^{\rm (II)} = \mathcal{L}_0 + \mathcal{L}_{\rm other} \;,
\end{equation}
where $\mathcal{L}_{\rm other}$ is the same as in the heterotic case \cite{Gross:1986mw}. As the 
lowest-order solutions (obtained from $\mathcal{L}_0$) all satisfy property (\ref{modrie0}), we 
conclude that higher-derivative term $\mathcal{L}_{\rm other}$ should be irrelevant in calculations 
of solutions and corresponding central charges, which stay $\alpha'$-uncorrected. 

This is a well-known fact obtained by other means in the literature. We have shown here how it can 
be understood as a simple consequence of the form of  the 10-dimensional tree-level effective 
actions of type-II theories.

\section{\label{sec:concl}Conclusion}

We have shown on several examples of string backgrounds containing AdS$_3$ factor how one 
can calculate in $\alpha'$-exact manner BPS and non-BPS solutions and corresponding conformal 
central charges from the \emph{complete} tree-level effective action (by taking into account all 
higher-derivative terms). Let  us here discuss some of the important issues and outcomes of our 
analysis, which are common to all examples: 
\begin{enumerate}
\item
Though our solutions were obtained from the reduced Lagrangian (\ref{redl}),
%\footnote{More 
%precisely, we used dual Lagrangian (\ref{tlred}), which is equivalent to (\ref{redl}).} 
from the fact that 
they all satisfy property (\ref{modrie0}) it follows that they are also solutions of the supersymmetric
Lagrangian (\ref{susyl}) (obtained in $D=10$ by $\mathcal{N}=1$ supersymmetry completion from 
Chern-Simons term). Agreement of our "macroscopic" results for central charges agree with those 
obtained "microscopically" shows that Chern-Simons terms are solely responsible for 
$\alpha'$-corrections. Now, this is not surprising for supersymmetric (BPS) solutions (\ref{solf6d2c}), 
(\ref{solf5d3c}) and (\ref{solf9d3c}), because it was shown in effective AdS$_3$ analyses 
\cite{Kraus:2006wn,David:2007ak,Kaura:2008us} that this is generally valid if supersymmetry is 
present in effective 3-dimensional theory. \emph{What is new here is that non-supersymmetric 
example (\ref{solfn5d3c}) shows that this happens also in cases where 10-dimensional 
supersymmetry is completely broken in effective 3-dimensional theory.} These examples suggest 
possible extension of the results from \cite{Kraus:2006wn,David:2007ak,Kaura:2008us}  to more 
general (non-supersymmetric) situations.
\item
This immediately leads us to the question why the part of the tree-level effective Lagrangian 
denoted $\mathcal{L}_{\rm other}$ in (\ref{10dl}) is not giving any contribution to central charges
(and, as we are suggesting, neither to the solutions). This part starts at $\alpha'^3$ (8-derivative) 
order and contains the (in)famous $\zeta(3)RRRR$ terms. Contrary to the 
$\Delta\mathcal{L}_{\rm CS}$ term, the structure of $\mathcal{L}_{\rm other}$ is grossly unknown, 
with only 4-point sector being completely known \cite{Gross:1986mw}. It was shown in 
\cite{Gross:1986mw} that this 4-point sector can be written in simple form 
$\zeta(3)\overline{R}\overline{R}\overline{R}\overline{R}$, where $\overline{R}$ stands for torsional 
Riemann tensor obtained from the modified ("torsional") connection (\ref{modcon}) (and written 
explicitly in (\ref{modrie})). This has stimulated authors of \cite{Kehagias:1997cq} to conjecture
that the whole $\mathcal{L}_{\rm other}$ can be written in such way (by using 
$\overline{R}_{MNPQ}$ only). Indeed, the most recent calculations of some 5-point terms 
($\zeta(3)RRRHH$ and $\zeta(3)R(\Delta H)(\Delta H)HH$) are in accord with this conjecture 
\cite{Richards:2008sa}. What is important for us here is \emph{if this conjecture 
is correct, it would immediately imply that term $\mathcal{L}_{\rm other}$ does not contribute to 
our solutions and central charges} (a proof is the same as in the case of 
$\Delta\mathcal{L}_{\rm CS}$). That would mean that we have found solutions of the \emph{full}
tree-level effective action(s). Of course, we can turn the argument around and claim that
agreement of our results (for central charges) with the microscopic calculations argues in
favor of the conjecture. However, we should be careful in making strong statements because of
the following reasons:
  \begin{enumerate}
  \item
It would be enough for our purposes that every monomial in $\mathcal{L}_{\rm other}$ is bilinear 
in $\overline{R}_{MNPQ}$. This weaker version of the conjecture (appearing already in 
\cite{Metsaev:1987zx}) would also imply that $\mathcal{L}_{\rm other}$ is irrelevant for our 
results.
  \item
Beside $\overline{R}_{MNPQ}=0$, our solutions satisfy other common properties, e.g., $R=0$
(vanishing of 10-dimensional Ricci scalar), $\overline{H}^2=0$, and on top of it all covariant 
derivatives are zero. So, adding to effective action terms which contain covariant derivative or 
bilinear in $R$,  $\overline{H}^2$ would not change our results, and so our analysis does not put 
any constraint on them. To clear this issue, we have to find examples in which we have 
$\overline{R}_{MNPQ}=0$ but not these other properties.
  \end{enumerate}
\item
Reduced Lagrangian $\widetilde{\mathcal{L}}_{\rm red}$ is of 4-derivative type. Our analyses 
offers direct explanation (in 10-dimensional set-up) why terms with six and more derivatives in 
10-dimensional string effective actions are irrelevant for calculations considered here and in 
\cite{Prester:2008iu}.
\item
All our solutions have the form of $\alpha'$-uncorrected solutions when we used magnetic
charges calculated in near-horizon region from 3-form $\overline{H}$. This is exactly what is 
obtained in sigma model calculations, despite the fact that those two methods are typically 
using different field-redefinition schemes (for example, this agreement is not manifestly 
present in corresponding black hole near-horizon analyses 
\cite{Prester:2008iu,Bellucci:2009nn}). 
\item
We have shown in section \ref{ssec:typeII} that our strategy can be trivially extended to NS-NS backgrounds in type II string theories. Now, structure of dualities connecting heterotic and type II 
string theories suggests that the same strategy could be further extended to Ramond-Ramond backgrounds in type II theories. Indeed, it has been proposed 
\cite{Rajaraman:2005up,Paulos:2008tn} that R-R 5-form field couples to the gravity at 8-derivative 
order exclusively through a relation similar with Eq. \ref{modrie} (see Eq. (2.10)-(2.11) from 
\cite{Paulos:2008tn}). If this proposal is correct, it could be used to argue for the vanishing of corresponding corrections to particular backgrounds, using the same logic which we applied here 
and in \cite{Prester:2008iu}.
\end{enumerate}

\begin{acknowledgments}
We thank Loriano Bonora and Gabriel Lopes Cardoso for stimulating discussions, and especially to
Ashoke Sen for invaluable explanations on definition of charges. This work was 
supported by the Croatian Ministry of Science, Education and Sport under the contract No.\ 119-0982930-1016, and by Alexander von Humboldt Foundation.
\end{acknowledgments}

\end{document}